\documentclass[12pt]{iopart}

\usepackage{epsfig}
\usepackage{graphicx}

\begin{document}

\title{Dressed molecules in an optical lattice}

\author{K. B. Gubbels, D. B. M. Dickerscheid, and H. T. C. Stoof}

\address{Institute for Theoretical Physics, Utrecht University,
Leuvenlaan 4, 3584 CE Utrecht, The Netherlands}
\ead{K.Gubbels@phys.uu.nl}

\begin{abstract}
We present the theory of an atomic gas in an optical lattice near
a Feshbach resonance. We derive from first principles a
generalized Hubbard model, that incorporates all the relevant
two-body physics exactly, except for the background atom-atom
scattering. For most atoms the background interactions are
negligible, but this is not true for ${}^6$Li, which has an
exceptionally large background scattering length near the
experimentally relevant Feshbach resonance at 834 G. Therefore, we
show how to include background atom-atom scattering by solving the
on-site two-body Feshbach problem exactly. We apply the obtained
solution to ${}^6$Li and find that the background interactions
indeed have a significant effect in this case.
\end{abstract}

 \submitto{\NJP}

\section{Introduction.}
The field of ultracold atoms has recently seen many exciting
developments in which strong interactions play a crucial role.
Examples are the observation of the quantum phase transition
between the superfluid and the Mott-insulator phase of a Bose gas
in an optical lattice \cite{quphtr}, the observation of the
BEC-BCS crossover in a Fermi gas
\cite{cross1,cross2,cross3,cross4,cross5,cross6}, and the
observation of fermionization in an one-dimensional Bose gas
\cite{fermionization1,fermionization2}. To a large extent these
developments are due to the successful implementation of two new
experimental techniques, namely the use of optical lattices
\cite{optlat} and the application of Feshbach resonances in the
collision of two atoms \cite{feshstwalley, feshstoof}. Both
techniques have contributed greatly to the unprecendented control
over the relevant physical parameters of an atomic gas.

In this article, we focus on the combination of these two
important techniques and present the microscopic theory of an
atomic gas in an optical lattice near a Feshbach resonance. This
problem was already studied in ref. \cite{feshopt}, where it was
argued that an optical lattice could be very well suited to
overcome the experimental difficulties in studying the quantum
phase transition that occurs in an atomic Bose gas near a Feshbach
resonance \cite{ACMC1,ACMC2}. The observation of this quantum
Ising transition between a phase with only a molecular condensate
and a phase with both an atomic and a molecular condensate is
complicated by the fast vibrational relaxation of the Feshbach
molecules due to collisions, which appears to prevent the creation
of a molecular condensate \cite{nomoco}. In an optical lattice
with low filling fractions, collisions can essentially be
neglected and this problem is expected to be much less severe.
Furthermore, in a subsequent study, a resonantly-interacting
Bose-Fermi mixture in an optical lattice was considered, which led
to the prediction of another XY-like quantum phase transition,
associated with the Bose-Einstein condensation of the bosons in
the mixture \cite{mixt}.

In both studies \cite{feshopt,mixt}, a generalized Hubbard model
was obtained by incorporating the relevant two-body physics
exactly. As a result, simple mean-field techniques could be used
to accurately describe the many-body physics of the atomic gas
near the Feshbach resonance in the optical lattice. This approach
led to some discussion \cite{comment,reply}, which motivated us to
rederive in this article the proposed generalized Hubbard model
from first principles by means of a field-theoretical calculation.
In this manner, the validity of the original approach is put on an
even more rigorous basis.

Another aspect that we discuss in this article, is the effect of
background atom-atom scattering on the Feshbach physics. This
effect was not incorporated in the previously mentioned studies,
since only systems with a small background scattering length were
considered. However, for ${}^6$Li, which has an exceptionally
large background scattering length near the experimentally
relevant Feshbach resonance at 834 G, the neglect of the
background interactions is no longer allowed. In this article, we
therefore show how to include the background atom-atom scattering
by exactly solving the two-body Feshbach problem on a single site
of the optical lattice. The exact knowledge of the two-body
physics on a single site can then be directly incorporated into a
generalized Hubbard model that describes the many-body physics of
${}^{6}$Li in an optical lattice near a Feshbach resonance, in a
similar manner as recently achieved for ${}^{40}$K
\cite{koetsier}.

\section{Effective atom-molecule theory.} \label{fieldtheory}
In this section, we give an {\it ab initio} field-theoretical
derivation of the effective atom-molecule theory that describes an
atomic gas in an optical lattice near a Feshbach resonance. The
theory is formulated in the tight-binding limit and incorporates
all the relevant two-body physics exactly. If desired, this rather
technical section can be omitted in a first reading of the
article. To facilitate this, section \ref{twobodphys} has been
written in such a way that it stands on its own. There, we show
how to solve analytically the two-channel Feshbach problem for two
atoms on a single site including background atom-atom scattering.
We also apply the obtained theory to ${}^6$Li.

\subsection{Effective action.}
The quantity of interest in a quantum-field theory is the
generating functional $Z$ of all the Green's functions. This
functional determines all the possible correlation functions of
the system. Specifically, let us consider the field theory for an
atom-molecule gas that is described by the action $S[\psi_{\rm
a}^{*},\psi_{\rm a}^{\phantom *}, \psi_{\rm m}^{*},\psi_{\rm
m}^{\phantom *}]$, with $\psi_{\rm a}$ and $\psi_{\rm m}$ being
the atomic and molecular fields, respectively. The generating
functional in imaginary time is defined by
\begin{eqnarray}
&&Z[J_{\rm a}^{*}, J_{\rm a}^{\phantom *}, J_{\rm m}^{*}, J_{\rm
m}^{\phantom *}] \nonumber \\ && = \int d[\psi_{\rm a}^{*}]~
d[\psi_{\rm a}^{\phantom *}] ~d[\psi_{\rm m}^{*}]~d[\psi_{\rm
m}^{\phantom *}] \exp{\left\{ -\frac{1}{\hbar} S[\psi_{\rm
a}^{*},\psi_{\rm a}^{\phantom *}, \psi_{\rm m}^{*},\psi_{\rm
m}^{\phantom *}] + S_{J} \right\} },
\end{eqnarray}
where the source currents couple to the fields according to,
\begin{eqnarray}
S_{J} = &&  \int d\tau \int d{\bf x} \left[ \psi_{\rm a}^{*}({\bf
x},\tau) J_{\rm a}({\bf x},\tau) + J_{\rm a}^{*} ({\bf x},\tau)
\psi_{\rm a}^{\phantom *}({\bf x},\tau) \right.
\nonumber \\
&& \left. + \psi_{\rm m}^{*}({\bf x},\tau) J_{\rm m}({\bf x},\tau)
+ J_{\rm m}^{*} ({\bf x},\tau) \psi_{\rm m}({\bf x},\tau) \right].
\end{eqnarray}
By taking functional derivatives of $Z$ with respect to the
currents we can calculate all the correlation functions of the
theory.

Instead of working with $Z$ we usually prefer to work with the
generating functional $W$ of all the connected Green's functions,
which is related to $Z$ through $Z = \exp(W)$. The functional
derivatives of $W$ with respect to the currents are the
expectation values of the fields, i.e.,
\begin{eqnarray}
\frac{\delta W}{\delta J_{\rm a}({\bf x},\tau)} &=& \langle
\psi_{\rm a}^{*} ({\bf x},\tau)\rangle \equiv \phi_{\rm
a}^{*}({\bf x},\tau) \\
\frac{\delta W}{\delta J_{\rm m}({\bf x},\tau)} &=& \langle
\psi_{\rm m}^{*} ({\bf x},\tau)\rangle \equiv \phi_{\rm
m}^{*}({\bf x},\tau).
\end{eqnarray}
Similar equations hold for the expectation values of the complex
conjugated fields $\psi_{\rm a}({\bf x},\tau)$ and $\psi_{\rm
m}({\bf x},\tau)$. Instead of using $W$, which only depends on the
current sources, it is possible to define a functional $\Gamma$
that depends explicitly on the fields $\phi_{\rm a}({\bf x},\tau)$
and $\phi_{\rm m}({\bf x},\tau)$, and which is related to $W$ by
means of a Legendre transformation, i.e.,
\begin{eqnarray}
\Gamma[ \phi_{\rm a}^{*}, \phi_{\rm a}^{\phantom *}, \phi_{\rm
m}^{*}, \phi_{\rm m}^{\phantom *}]
 &=& - W[J_{\rm a}^{*}, J_{\rm a}^{\phantom *}, J_{\rm m}^{*}, J_{\rm m}^{\phantom *}]
\nonumber \\
&&+ \int d\tau~ \int d{\bf x}~ \left[ \phi_{\rm a}^{*}({\bf
x},\tau ) J_{\rm a}({\bf x},\tau ) + J_{\rm a}^{*} ({\bf x},\tau )
\phi_{\rm a}^{\phantom *}({\bf x},\tau ) \right.
\nonumber \\
&&+ \left. \phi_{\rm m}^{*}({\bf x},\tau ) J_{\rm m}({\bf x},\tau
) + J_{\rm m}^{*} ({\bf x},\tau) \phi_{\rm m}^{\phantom *}({\bf
x},\tau ) \right].
\end{eqnarray}
The reason for defining this functional is that $\Gamma$ is
related to the exact effective action of the system through
$S^{\rm eff} = - \hbar \Gamma$. Technically, $\Gamma$ is the
generating functional of all one-particle irreducible vertex
functions. In our case, the exact effective action for the
atom-molecule theory can be written as
\begin{eqnarray} \label{effact}
S^{\rm eff}[ \phi_{\rm a}^{*}, \phi_{\rm a}^{\phantom *},
\phi_{\rm m}^{*}, \phi_{\rm m}^{\phantom *}] &=& {\rm Tr} \left[ -
\phi_{\rm a}^{*} \hbar G^{-1}_{\rm a} \phi_{\rm a}^{\phantom *} -
\phi_{\rm m}^{*} \hbar G^{-1}_{\rm m} \phi_{\rm m}^{\phantom *}
\right. \nonumber \\ &&+ \left. g \left( \phi_{\rm m}^{*}
\phi_{\rm a}^{\phantom *} \phi_{\rm a}^{\phantom *} + \phi_{\rm
a}^{*} \phi_{\rm a}^{*} \phi_{\rm m}^{\phantom *} \right) + \ldots
\right],
\end{eqnarray}
where $G_{\rm a/m}$ is the exact propagator of the
atoms/molecules, and $g$ is the exact three-point vertex. The dots
denote all possible other one-particle irreducible vertices, which
turn out to be less relevant for our purposes.

\subsection{Microscopic action.}
Starting from the microscopic atom-molecule theory in an optical
lattice, it is our goal to derive an effective quantum field
theory, that contains the relevant two-body physics exactly. For
calculational convenience and to facilitate comparison with
previous work \cite{feshopt}, we consider the case of bosonic
atoms. The generalization to fermionic atoms is then
straightforward.

The total microscopic action $S$ describing resonantly-interacting
atoms in an optical lattice can be split up in three parts, namely
a purely atomic part $S_{\rm a}$, a purely molecular part $S_{\rm
m}$ and a coupling between atoms and molecules $S_{\rm am}$,
\begin{eqnarray}
S = S_{\rm a} + S_{\rm m} + S_{\rm am}.
\end{eqnarray}
The purely atomic part $S_{\rm a}$ is given by
\begin{eqnarray}
S_{\rm a} &=& \int_{0}^{\hbar \beta} d \tau \int d{\bf x}~
\psi_{\rm a}^{*} ({\bf x},\tau) \left( \hbar \partial_{\tau} -
\frac{\hbar^{2} \nabla^{2}}{2 m_{\rm a}} - \mu + V_{0} ({\bf x})
\right) \psi_{\rm a}({\bf x},\tau) ^{\phantom *} \nonumber \\ && +
\frac{1}{2} \int_{0}^{\hbar \beta} d \tau \int d{\bf x}~ \frac{4
\pi a_{\rm bg} \hbar^{2}}{m_{\rm a}} \psi^{*}_{\rm a}({\bf
x},\tau)
    \psi^{*}_{\rm a}({\bf x},\tau)
 \psi_{\rm a}({\bf x},\tau) ^{\phantom *} \! \!
    \psi_{\rm a}({\bf x},\tau) ^{\phantom *},
\end{eqnarray}
with $\mu$ the chemical potential, $m_{\rm a}$ the atomic mass,
$a_{\rm bg}$ the background scattering length and $V_{0} ({\bf
x})$ the external periodic potential due to the optical lattice.
In first instance, we neglect the background atom-atom scattering,
since we are primarily interested in the resonant interactions
between the atoms and the molecules. For most atoms, this is an
accurate approximation. Furthermore, the same approximation was
also used in previous work \cite{feshopt}, with which we
ultimately want to compare the following derivation. Finally, in
section \ref{twobodphys} we overcome this approximation and show
how to include the effect of background atom-atom scattering.

The purely molecular action $S_{\rm m}$ is given by
\begin{eqnarray}\label{defsm}
S_{\rm m} &=&
 \int_{0}^{\hbar \beta} d \tau
\int d{\bf x}~ \psi_{\rm m}^{*} ({\bf x},\tau) \left( \hbar
\partial_{\tau} - \frac{\hbar^{2} \nabla^{2}}{4 m_{\rm a}} + \delta_{\rm B} -
2 \mu +2 V_{0} ({\bf x})  \right) \psi_{\rm m}({\bf
x},\tau)^{\phantom *}. \nonumber \\
\end{eqnarray}
where $\delta_{\rm B}$ is the so-called bare detuning (see also
section \ref{twobodphys}). The action $S_{\rm am}$ corresponding
to the atom-molecule coupling that describes the formation of a
bare molecule from two atoms and vice versa, is given by
\begin{eqnarray}
S_{\rm am} &=& \int_{0}^{\hbar \beta} d \tau \int d {\bf x} \int
d{\bf x}'~ g ({\bf x} - {\bf x'}) \left\{ \psi_{\rm m}^{*} (({\bf
x} + {\bf x}') /2, \tau) \psi_{\rm a}({\bf x}',\tau)^{\phantom *}
\! \psi_{\rm a}({\bf x},\tau)^{\phantom *} \right. \nonumber \\ &&
\left. + \psi^{*}_{\rm a}({\bf x}',\tau) \psi^{*}_{\rm a}({\bf
x},\tau) \psi_{\rm m}^{\phantom *}  (({\bf x} + {\bf x}') /2,
\tau) \right\},
\end{eqnarray}
where $g({\bf x})$ denotes the atom-molecule coupling.

We can write the actions involving the atomic fields in a more
convenient way as
\begin{eqnarray}
S_{\rm a} + S_{\rm am} &=& -\frac{\hbar}{2} \int_{0}^{\hbar \beta}
d
\tau d \tau' \int d{\bf x} d{\bf x}'~ \nonumber \\
&& \times \Big( \psi_{\rm a}^{*} ({\bf x},\tau), \psi_{\rm
a}^{\phantom *} ({\bf x},\tau) \Big) {\bf G}^{-1}({\bf
x},\tau;{\bf x'},\tau') \left(\begin{array}{c} \psi_{\rm
a}^{\phantom *} ({\bf x}',\tau') \\
\psi_{\rm a}^{ *} ({\bf x}',\tau') \end{array} \right),
\end{eqnarray}
with ${\bf G}$ the $2 \times 2$ (Nambu space) Green's function
matrix, given by
\begin{equation}\label{defga}
{\bf G}^{-1} = {\bf G}_{{\rm a}}^{-1} - {\bf \Sigma}= {\bf
G}_{{\rm a}}^{-1} \left( 1 - {\bf G}_{{\rm a}} {\bf \Sigma}
\right) .
\end{equation}
Here, the atomic Green's function matrix ${\bf G}_{{\rm a}}^{-1}$
has the following form
\begin{equation}
{\bf G}^{-1}_{{\rm a}} = \left[
\begin{array}{cc}
G^{-1}_{{\rm a}}({\bf x}, \tau ; {\bf x}', \tau') & 0 \\
0 & G^{-1}_{{\rm a}}({\bf x}', \tau' ; {\bf x}, \tau)
\end{array}
\right],
\end{equation}
where the atomic zeroth-order Green's function $G_{{\rm a}}$
satisfies
\begin{eqnarray}\label{zerogf}
\left\{ \hbar \partial_{\tau} - \frac{\hbar^{2} \nabla^{2}}{2
m_{\rm a}} + V_{0} ({\bf x}) - \mu \right\} G_{{\rm a}}({\bf
x},\tau;{\bf x}',\tau') = -\hbar \delta(\tau - \tau') \delta({\bf
x} - {\bf x}').
\end{eqnarray}
The self-energy matrix $\hbar {\bf \Sigma}$ is given by
\begin{eqnarray}
\hbar {\bf \Sigma} &=& \left[
\begin{array}{cc}
0 & 2 g({\bf x} - {\bf x}') \psi_{\rm m}(({\bf x} + {\bf x}'
)/2,\tau)^{\phantom *}  \\
2 g({\bf x} - {\bf x}') \psi_{\rm m}^{*}(({\bf x} + {\bf
x}')/2,\tau) & 0
\end{array}
\right] \nonumber \\
&& \times \delta (\tau-\tau').
\end{eqnarray}

\begin{figure}[t]
\centering
\includegraphics[width=0.6\textwidth]{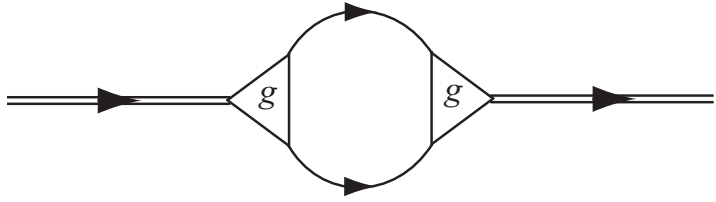}
\caption{Diagrammatic representation of the term $-{\rm Tr} \left[
{\bf G}_{{\rm a}} {\bf \Sigma} {\bf G}_{{\rm a}} {\bf \Sigma }
\right]/4$, which expresses that a bare molecule can break up into
two atoms and recombine again. The coupling strength between the
atoms and the molecule is given by $g$. This diagram can be
interpreted as a self-energy of the bare molecules.
\label{feynman}}
\end{figure}

From the microscopic atom-molecule action we can calculate the
grand-canonical partition function $Z$ as the functional integral
\begin{eqnarray}
Z = \int d[\psi_{\rm a}^{*}] d[\psi_{\rm a}^{\phantom *}]
d[\psi^{*}_{\rm m}] d[\psi^{\phantom *}_{\rm m}] \exp{\left\{
-\frac{1}{\hbar} S[\psi_{\rm a}^{*},\psi_{\rm a}^{\phantom *},
\psi_{\rm m}^{*},\psi_{\rm m}^{\phantom *}] \right\}}.
\end{eqnarray}
Since the integral is Gaussian in the atomic fields, we can
perform this integral exactly, giving
\begin{eqnarray} \label{effZ}
Z &=& \int d[\psi^{*}_{\rm m}] d[\psi_{\rm m}]
\exp{\left\{ -\frac{1}{2}\Tr[ \ln (-{\bf G}^{-1})] - \frac{1}{\hbar}S_{\rm m} \right\}} \nonumber \\
&=& \int d[\psi^{*}_{{\rm m}}] d[\psi^{\phantom *}_{{\rm m}}]
\exp{\left\{-\frac{1}{2}\Tr[ \ln (-{\bf G}_{\rm a}^{-1})]
-\frac{1}{2}\Tr[ \ln ({\bf 1}-{\bf G}_{\rm a}{\bf \Sigma})] -
\frac{1}{\hbar}S_{\rm
m}\right\}}, \nonumber \\
\end{eqnarray}
where the traces are taken over a $2 \times 2$ matrix structure,
coordinate space, and imaginary time. By integrating out the
atoms, we have obtained an effective action for the bare molecules
that consists of three terms. The first term represents the
contribution from the ideal atomic gas, which does not depend on
the bare molecular fields. The second term can be brought in a
more illuminating form by expanding the logarithm. Because of the
trace, only even powers in the expansion give nonzero results,
yielding
\begin{equation} \label{expansion}
\frac{1}{2}\Tr[ \ln (1-{\bf G}_{\rm a}{\bf
\Sigma})]=-\sum_{n=2,4,...} \frac{1}{2 n} ({\bf G}_{{\rm a}}{\bf
\Sigma})^n.
\end{equation}
The first nonzero term is of second order, namely $-{\rm Tr}
\left[ {\bf G}_{{\rm a}} {\bf \Sigma} {\bf G}_{{\rm a}} {\bf
\Sigma } \right]/4$, and expresses that a molecule can break up
into two atoms and recombine again. This term can be interpreted
as corresponding to the self-energy of the bare molecules and a
diagrammatic picture of the term is shown in figure \ref{feynman}.
The higher-order terms in equation (\ref{expansion}) pertain to
interactions between the bare molecules.

\subsection{Molecular self-energy.}
Next, we evaluate the molecular self-energy term, $-{\rm Tr}
\left[ {\bf G}_{{\rm a}} {\bf \Sigma} {\bf G}_{{\rm a}} {\bf
\Sigma } \right]/4$, by explicitly carrying out the trace. As a
first step, we introduce center-of-mass coordinates ${\bf R} =
({\bf x}' + {\bf x}'')/2$, ${\bf R}' = ({\bf x} + {\bf x}''')/2$
and relative coordinates ${\bf r} = {\bf x}' - {\bf x}''$, ${\bf
r}' = {\bf x}''' - {\bf x}$. Writing out the trace in these
coordinates yields
\begin{eqnarray}\label{diageq}
&-&\frac{1}{4}{\rm Tr} \left[ {\bf G}_{{\rm a}} {\bf \Sigma} {\bf
G}_{{\rm a}} {\bf \Sigma } \right] = -\frac{2}{\hbar^2}
\int_{0}^{\hbar \beta} d \tau ~d \tau' \int d{\bf R}~ d{\bf R}'~
d{\bf r}~ d{\bf r}'~
g({\bf r}) g({\bf r}')\psi_{\rm m}^{*} ({\bf R},\tau)  \nonumber \\
&&\times \psi_{\rm m} ({\bf R}',\tau')^{\phantom *} G_{{\rm
a}}({\bf R}' + {\bf r}' /2,\tau; {\bf R} + {\bf r}/2,\tau')
G_{{\rm a}}({\bf R}' - {\bf r}' /2,\tau; {\bf R} -
{\bf r}/2,\tau') \nonumber \\
&=&- \frac{2 g^{2}}{\hbar^2} \int_{0}^{\hbar \beta} d \tau ~d
\tau' \int d{\bf
R}~ d{\bf R}' \psi_{\rm m}^{*} ({\bf R},\tau)  \psi_{\rm m} ({\bf R}',\tau')^{\phantom *} \nonumber \\
&& \times G_{{\rm a}}({\bf R}',\tau; {\bf R},\tau') G_{{\rm
a}}({\bf R}',\tau;{\bf R},\tau'),
\end{eqnarray}
where in the last line we made use of the usual pseudopotential
approximation $g({\bf r}) = g \delta({\bf r})$.


In order to proceed, we rewrite the atomic Green's functions by
performing an expansion for both their spacial and temporal
arguments. For the expansion of the space dependence, we use the
complete set of atomic Bloch wave functions $\chi^{\rm a}_{\bf
n,k}$, which are solutions to the Schr\"{o}dinger equation for a
particle in a periodic potential
\begin{eqnarray}
\left[ -\frac{\hbar^{2} \nabla^{2}}{2 m_{\rm a}} + V_{0} ({\bf x})
\right] \chi^{\rm a}_{\bf n,k} ({\bf x}) = \epsilon_{\bf n,k}
\chi^{\rm a}_{\bf n,k} ({\bf x}).
\end{eqnarray}
Here, the exact Bloch wave functions are given by
\begin{equation}
\chi^{\rm a}_{\bf n,k}({\bf x}) =\frac{1}{\sqrt{N_{\rm s}}} \sum_i
e^{i {\bf k\cdot x}_i}w^{\rm a}_{\bf n} ({\bf x-x}_i),
\end{equation}
with $w^{\rm a}_{\bf n} ({\bf x})$ the so-called (atomic) Wannier
functions, ${\bf x}_i$ the location of the optical lattice sites,
$N_{\rm s}$ the total number of sites, ${\bf k}$ the lattice
momentum, and ${\bf n}$ a set of quantum numbers for the various
Bloch bands in the optical lattice. The Fourier expansion in
imaginary time is performed by using the well-known Matsubara
modes \cite{fetter}. Combining the two expansions, we obtain
\begin{equation} \label{gffourier}
G_{{\rm a}} ({\bf x},\tau; {\bf x}',\tau') = \frac{1}{\hbar \beta}
\sum_{n} \sum_{{\bf n,k}} G_{{\rm a}} ({\bf n,k};i \omega_{n})
\chi_{\bf n,k} ({\bf x}) \chi^{*}_{\bf n,k} ({\bf x}') e^{-i
\omega_{n} (\tau - \tau')}.
\end{equation}
with $\omega_n$ the Matsubara frequencies and $G_{{\rm a}}({\bf
n,k};i \omega_{n})$ given by
\begin{equation}\label{gkdef}
G_{{\rm a}}({\bf n,k};i \omega_{n})=\frac{-\hbar}{- i \hbar
\omega_{n} + \epsilon_{\bf n,k} - \mu }.
\end{equation}
Next, we also expand the space and imaginary-time arguments of the
molecular fields $\psi_{\rm m}({\bf x}, \tau)$. For the temporal
expansion we use again the Matsubara modes and for the spacial
expansion we use the (molecular) Wannier functions $w^{\rm m}_{\bf
n}({\bf x})$. As a result,
\begin{equation} \label{wanexp}
\psi_{\rm m} ({\bf x},\tau) = \frac{1}{\sqrt{\hbar \beta}}
\sum_{{n}} \sum_{{\bf n},i} b_{{\bf n},i}(\omega_{n}) w^{\rm
m}_{{\bf n}}({\bf x} - {\bf x}_{i})
 e^{-i \omega_{n} \tau}.
\end{equation}
with the complex conjugated version for $\psi_m^*.$ The operators
$b^{\phantom \dagger}_{{\bf n},i}$ annihilate a molecule with a
set of quantum numbers ${\bf n}$ on site $i$ of the lattice.

Substituting equations (\ref{gffourier}) and (\ref{wanexp}) into
(\ref{diageq}), yields
\begin{eqnarray}\label{fridaymorning1}
&&-\frac{1}{4}{\rm Tr} \left[ {{\bf G}_{{\rm a}} {\bf \Sigma} {\bf
G}_{{\rm a}} {\bf \Sigma} } \right] =-\frac{2 g^{2}}{\hbar^5
\beta^3} \sum_{{n},{n'}} \sum_{m,m'} \sum_{i,j} \sum_{{\bf m,m'}}
\sum_{{\bf n},{\bf n'}}\sum_{{\bf k},{\bf k}'} \int_{0}^{\hbar
\beta} d \tau ~d \tau' \int d{\bf R}~ d{\bf R}'
\nonumber \\
&& \hspace{0.5in} \times  b^{*}_{{\bf m},i}  (\omega_{n})
b^{\phantom *}_{{\bf m' },j}  (\omega_{n'})
 e^{i \omega_{n} \tau}
 e^{-i \omega_{n'} \tau'}
e^{-i (\omega_{m} + \omega_{m'}) (\tau - \tau')}  \nonumber \\
&& \hspace{0.5in} \times w^{{\rm m}*}_{\bf m}({\bf R} - {\bf
x}_{i}) w^{\rm m}_{\bf m'}({\bf R}' - {\bf x}_{j}) G_{{\rm a}}
({\bf n,k};i \omega_{m}) G_{{\rm a}} ({\bf n}',{\bf k'};i
\omega_{m'})
\nonumber \\
&& \hspace{0.5in} \times \chi^{\rm a}_{\bf n,k} ({\bf R})
\chi^{{\rm a}*}_{\bf n,k} ({\bf R}') \chi^{\rm a}_{{\bf n',k'}}
({\bf R}) \chi^{{\rm a} *}_{{\bf n',k'}} ({\bf R}') .
\end{eqnarray}
The integrals over the imaginary time variables $\tau$ and $\tau'$
can now easily be performed, resulting in the conditions
$\omega_{n} = \omega_{n'} = \omega_{m} + \omega_{m'}$.
Furthermore, if the optical lattice is deep, it can at every site
be well approximated by an harmonic potential, characterized by
the frequency $\omega$. Then, we are also allowed to use the
so-called tight-binding limit, in which the Wannier functions
$w_{\bf n}$ are replaced by the tight-binding functions $\phi_{\bf
n}$, i.e.
\begin{equation}\label{tightbbb}
w_{\bf n} ({\bf R} - {\bf x}_{i}) \simeq \phi_{\bf n} ({\bf R} -
{\bf x}_{i}).
\end{equation}
Here, $\phi_{\bf n} ({\bf R} )$ are the eigenstates of the
three-dimensional isotropic harmonic oscillator potential given in
spherical coordinates by
\begin{eqnarray}
\phi_{\bf n} ({\bf R}) \equiv \phi_{n,\ell,m} ({\bf R}) &=&
\sqrt{\frac{2}{l^{3}}} \left( \begin{array}{c} n + \ell + 1/2 \\ n
\end{array} \right)^{-1/2} \frac{1}{\sqrt{(\ell+1/2)!}} \nonumber
\\ && \times e^{-R^{2}/2l^{2}} L_{n}^{(1/2 + \ell)}
((R/l)^{2})~(R/l)^{\ell}~Y_{\ell m} (\theta,\phi),
\end{eqnarray}
where $L_{n}^{(1/2 + \ell)}(X)$ are the generalized Laguerre
polynomials and $Y_{\ell,m} (\theta,\phi)$ the spherical
harmonics. For the atoms, the harmonic oscillator length $l$ is
given by $l_{\rm a}= \sqrt{\hbar / m_{\rm a} \omega}$; for the
molecules, the harmonic oscillator length $l$ is given by $l_{\rm
a}/\sqrt{2}$. Furthermore, in the extreme tight-binding limit the
energies $\epsilon_{\bf n,k}$ in equation (\ref{gkdef}) are given
by $\epsilon_{\bf n,k} = \epsilon_{\bf n} = (2 n + \ell +
3/2)\hbar \omega$ and only depend on the radial and angular
momentum quantum numbers $n$ and $\ell$, respectively. Similarly,
we have that $ G_{{\rm a}} ({\bf n},{\bf k};i \omega_{m})=G_{{\rm
a}} ({\bf n};i \omega_{m})$. Also note that in the tight-binding
limit, the overlap for wave functions on different lattice sites
is negligible. As a result, we can consider only the terms on the
right-hand side of equation (\ref{fridaymorning1}) that have all
atomic and molecular Wannier functions centered around the same
site ${\bf x}_{i}$. Furthermore, because of translational
invariance of the lattice, we can set ${\bf x}_{i}$ to zero in the
integrant. As a result, the right-hand side of equation
(\ref{fridaymorning1}) does not depend on ${\bf k,k'}$ anymore,
and the sum over the lattice momenta yields the factor $N_{\rm
s}^2$.

Next, we evaluate the sum over Matsubara frequencies, keeping the
external frequency $\omega_{m} + \omega_{m'} = \omega_{n}$ of the
molecules fixed. First, we rewrite the product of Green's
functions as
\begin{equation}
G_{{\rm a}} ({\bf n},i \omega_{m}) G_{{\rm a}} ({\bf n'},i
\omega_{m'}) = -\hbar \frac{ G_{{\rm a}} ({\bf n},i \omega_{m}) +
G_{{\rm a}} ({\bf n'},i \omega_{m'}) } { -i \hbar (\omega_{m} +
\omega_{m'}) + \epsilon_{\bf n} + \epsilon_{\bf n'} - 2 \mu }.
\end{equation}
Subsequently, we perform the sum and take the two-body limit,
which amounts to taking the atomic distribution functions,
obtained after the Matsubara summation, equal to zero
\cite{fetter}. We find
\begin{eqnarray}\label{fridayafternoon}
&&-\frac{1}{4}{\rm Tr} \left[ {{\bf G}_{{\rm a}} {\bf \Sigma} {\bf
G}_{{\rm a}} {\bf \Sigma} } \right] = - \frac{2 g^{2}}{\hbar}
\sum_{i} \sum_{n} \sum_{{\bf m},{\bf m'}}  \sum_{{\bf n},{\bf n}'}
\int~d{\bf R}~ d{\bf R}'~ b^{*}_{{\bf m},i}  (\omega_{n})
b^{\phantom *}_{{\bf m'},i}  (\omega_{n}) \nonumber \\
&& \times \frac{ \phi^{{\rm m}*}_{{\bf m}}({\bf R} ) \phi^{{\rm
m}\phantom *}_{{\bf m'}}({\bf R}' ) \phi^{\rm a}_{\bf n} ({\bf R})
\phi^{{\rm a}*}_{\bf n} ({\bf R}') \phi^{\rm a}_{{\bf n}'} ({\bf
R}) \phi^{{\rm a}*}_{{\bf n}'} ({\bf R}') }{-i\hbar \omega_{n} +
\epsilon_{\bf n} + \epsilon_{\bf n'} - 2 \mu},
\end{eqnarray}
where $\phi^{\rm m/a}_{\bf n}({\bf R})$ denotes the
molecular/atomic tight-binding function. For systems of interest
to us, the molecules occupy only the lowest Bloch band, i.e., we
take the band indices ${\bf m}$ and ${\bf m'}$ of the molecular
fields in the last expression equal to zero. The tight-binding
function $\phi^{\rm m}_{{\bf 0}}$ is given by $\phi^{{\rm m}}_{\bf
0} ({\bf R}) = \left(2/\pi l_{\rm a}^{2} \right)^{3/4}
e^{-(R/l_{\rm a})^{2}}$.

Next, we need to evaluate the integrals over ${\bf R}$ and ${\bf
R}'$ of the form
\begin{eqnarray}\label{triprod}
&&\int d{\bf R} \phi^{{\rm m}*}_{\bf 0}({\bf R} )\phi^{\rm a}_{\bf
n} ({\bf R}) \phi^{\rm a}_{{\bf n}'} ({\bf R}) = 2
\left(\frac{2}{\pi l_{\rm a}^{2}} \right)^{3/4} \nonumber \\
&& \times
\int dR~d\theta~d\phi~ R^{2 + \ell + \ell'} \sin{\theta}~ e^{-2
R^{2}}\frac{L_{n}^{(1/2 + \ell)} (R^{2})}{\sqrt{(\ell + 1/2)!}}
\frac{ L_{n'}^{(1/2 + \ell')} (R^{2}) }{\sqrt{(\ell' + 1/2)!}}
\nonumber \\
&&\times   Y_{\ell m} (\theta, \phi) Y_{\ell' m'} (\theta, \phi)
\left(
\begin{array}{c} n + \ell + 1/2 \\ n \end{array} \right)^{-1/2}
\left( \begin{array}{c} n' + \ell' + 1/2 \\ n' \end{array}
\right)^{-1/2},
\end{eqnarray}
where on the right-hand side of the above equation the integration
variable $R$ is made dimensionless using the length $l_{\rm a}$.
The integrations over the angles can be directly evaluated using
the orthonormality relations for the spherical harmonics. To
evaluate the remaining integral over $R$, we make use of the
following relation \cite{gradrys}
\begin{eqnarray}
&& \int_0^{\infty} dX~ e^{-2 X^{2}} X^{2 + 2\ell} L^{(1/2 +
\ell)}_{n}
(X^{2}) L^{(1/2 + \ell)}_{n'}(X^{2})\nonumber \\
&&\hspace{1in} = \frac{1}{2} \int_0^{\infty} dy~ e^{-2 y} y^{1/2 +
\ell}
 L^{(1/2 + \ell)}_{n} (y)
 L^{(1/2 + \ell)}_{n'} (y) \nonumber \\
&& \hspace{1in} =  \frac{\Gamma(n+n' + \ell + 3/2)}{2 n! n'!}
\frac{1}{2^{n+n'+\ell + 3/2}}.
\end{eqnarray}
Using this, we find that equation (\ref{triprod}) can be written
as
\begin{eqnarray} \label{integral}
&&\int d{\bf R} \phi^{{\rm m}*}_{\bf 0}({\bf R} )\phi^{\rm a}_{\bf
n} ({\bf R}) \phi^{\rm a}_{{\bf n}'} ({\bf R}) = (-1)^m
\delta_{-m,m'} \left( \frac{2}{\pi l_{\rm a}^{2}} \right)^{3/4}
 \frac{\Gamma(n+n' + \ell + 3/2)}{ n! n'!}
\nonumber \\ && \times \frac{1}{2^{n+n'+\ell + 3/2}}
\frac{1}{(\ell + 1/2)!} \left( \begin{array}{c} n + \ell + 1/2 \\
n
\end{array} \right)^{-1/2} \left( \begin{array}{c} n' + \ell + 1/2
\\ n' \end{array} \right)^{-1/2}.
\end{eqnarray}

Finally, we have to perform the sum over the quantum numbers of
equation (\ref{integral}). We first partially evaluate the
resulting sum over the quantum numbers ${\bf n}$ and ${\bf n}'$ by
summing out all the contributions that have the same energy
$\epsilon_{q} \equiv \epsilon_{q,0} = \epsilon_{n,\ell}
+\epsilon_{n',\ell}- 3 \hbar \omega/2$. This yields
\begin{eqnarray} \label{diffsum}
 \sum_{{\bf n},{\bf n'}} \left[\int d{\bf R} \phi^{{\rm m}*}_{\bf
0}({\bf R} )\phi^{\rm a}_{\bf n} ({\bf R}) \phi^{\rm a}_{{\bf n}'}
({\bf R}) \right]^{2}\delta_{q,n+n'+\ell} &=& \phi^{{\rm aa} *
}_{q} ({\bf 0}) \phi^{{\rm aa} }_{q} ({\bf 0}),
\end{eqnarray}
where the relative wavefunctions of two particles in a harmonic
potential $\phi^{{\rm aa}}_{q} ({\bf r})$ will be introduced in
section \ref{twobodphys} and are given by equation (\ref{relwf}).
The reason for expressing the sum in terms of the relative
wavefunctions $\phi^{{\rm aa}}_{q} ({\bf r})$ is to reveal the
complete agreement between the field-theoretical calculation in
this section and the two-body calculations in section
\ref{twobodphys}.

For the second order term corresponding to the molecular
self-energy we now obtain
\begin{eqnarray} \label{tracef}
&&-\frac{1}{4}{\rm Tr} \left[ {{\bf G}_{{\rm a}} {\bf \Sigma} {\bf
G}_{{\rm a}} {\bf \Sigma} } \right] = - 2 g^{2} \sum_{i,n}
b^{*}_{{\bf 0},i}  (\omega_{n}) b^{\phantom *}_{{\bf 0},i}
(\omega_{n}) \nonumber
\\ && \times \sum_{q} \frac{ \phi^{{\rm aa}*}_{q}({\bf 0}) \phi^{\rm aa}_{q}({\bf 0}) }{-i\hbar \omega_{n} + \epsilon_{q} + 3 \hbar
\omega/2  - 2 \mu}.
\end{eqnarray}
This sum contains an ultraviolet divergence resulting from the use
of pseudopotentials. In section \ref{twobodphys}, we show how to
deal with this divergence by means of a renormalization procedure,
which leads to the final form of the molecular self-energy $\hbar
\Sigma_{\rm m}$
\begin{eqnarray} \label{sigmaf}
\hbar \Sigma_{\rm m} (i \hbar \omega_{n}) = g^{2} \frac{\Upsilon(
  i \hbar \omega_{n} - 3 \hbar \omega/2 + 2 \mu
)}{\sqrt{2} \pi l_{\rm a}^{3} \hbar \omega},
\end{eqnarray}
where $\Upsilon(z)$ is the ratio of two gamma functions
\begin{equation} \label{ups}
\Upsilon(z) \equiv \frac{\Gamma(-z/2 \hbar \omega + 3/4)}{
\Gamma(-z/2 \hbar \omega + 1/4)}.
\end{equation}
Note that the self-energy of equation (\ref{sigmaf}) is completely
equivalent to the self-energy obtained from a two-body calculation
in ref. \cite{feshopt}.

\subsection{Generalized Hubbard model}
So far, we have calculated the self-energy of the bare molecules,
but we have not yet taken into account the effects due to the
purely molecular action $S_{\rm m}$ and the purely atomic action
$S_{\rm a}$. This is what we do next.

The purely action $S_{\rm m}$, given by equation (\ref{defsm}),
can be rewritten by using equation (\ref{wanexp}), which yields
\begin{eqnarray}\label{quadmol}
S_{\rm m} &=& \int_{0}^{\hbar \beta} d\tau \int d{\bf x}~
\psi_{\rm m}^{*} ({\bf x},\tau) \left( \hbar \partial_{\tau} -
\frac{\hbar^{2} \nabla^{2}}{4 m_{\rm a}} + \delta_{\rm B} - 2 \mu
+ 2V_{0} ({\bf x}) \right) \psi_{\rm m}^{\phantom *} ({\bf
x},\tau) \nonumber \\ &=& \sum_{{n}} \sum_{i,j} \sum_{{\bf m},{\bf
m'}} b^{*}_{{\bf m},i} (\omega_{n}) b^{\phantom *}_{{\bf m'}, j}
(\omega_{n})\int d{\bf x}~ w^{{\rm m}*}_{{\bf m}}({\bf x} - {\bf
x}_{i}) \nonumber \\
&& \times  \left( - i \hbar \omega_{n}  - \frac{\hbar^{2}
\nabla^{2}}{4 m_{\rm a}} +2 V_{0} ({\bf x}) + \delta_{\rm B} - 2
\mu \right) w^{\rm m}_{{\bf m'}} ({\bf x} - {\bf x}_{j}),
\end{eqnarray}
From which we can immediately read off the hopping term $t_{\rm
m}$ and the on-site energy $\epsilon_{\rm m}$. For our purposes,
the important case is when the molecules are in the lowest band.
Then, we find for the hopping parameter $t_{\rm m}$
\begin{eqnarray}\label{hopmolec}
t_{{\rm m}} = - \int d{\bf x}~
 w^{{\rm m}*}_{\bf 0}({\bf x} - {\bf x}_{i})
\left[
 - \frac{\hbar^{2} \nabla^{2}}{4 m_{\rm a}}
+2 V_{0} ({\bf x}) \right] w^{{\rm m}}_{\bf 0} ({\bf x} - {\bf
x}_{j}),
\end{eqnarray}
where $i$ and $j$ are nearest-neighbouring sites. The on-site
energy $\epsilon_{\rm m}$ is given by
\begin{eqnarray}
\epsilon_{{\rm m}} =
 \int d{\bf x}~
 w^{{\rm m}*}_{\bf 0}({\bf x} - {\bf x}_{i})
\left[
 - \frac{\hbar^{2} \nabla^{2}}{4 m_{\rm a}}
+2 V_{0} ({\bf x}) \right] w^{{\rm m}}_{\bf 0} ({\bf x} - {\bf
x}_{i}).
\end{eqnarray}
As a result, the action $S_{\rm m}$ can be written as a lattice
action
\begin{eqnarray} \label{latticesm}
S_{\rm m} &=& \sum_{{n}}  \left\{ - t_{\rm m} \sum_{\langle i, j
\rangle } b^{*}_{i} (\omega_{n}) b^{\phantom *}_{j} (\omega_{n})
\right. \nonumber \\ && \left. + \sum_{i} \left( -i \hbar
\omega_{n} + \epsilon_{\rm m} + \delta_{\rm B} - 2 \mu \right)
b^{*}_{i} (\omega_{n}) b^{\phantom *}_{i} (\omega_{n}) \right\},
\end{eqnarray}
where $\langle i , j \rangle$ denotes the sum over nearest
neighbours and $\epsilon_{\rm m} = 3 \hbar \omega /2$ in the
tight-binding limit.

In exactly the same way, we can rewrite the atomic action $S_{\rm
a}$ for the lowest Bloch band, yielding
\begin{eqnarray}
S_{\rm a} &=& \sum_{{n}}  \left\{ - t_{\rm a} \sum_{\langle i, j
\rangle } a^{*}_{i} (\omega_{n}) a^{\phantom *}_{j}
(\omega_{n})\right. \nonumber \\ && \left. + \sum_{i} \left( -i
\hbar \omega_{n} + \epsilon_{\rm a} - \mu \right) a^{*}_{i}
(\omega_{n}) a^{\phantom *}_{i} (\omega_{n}) \right\},
\end{eqnarray}
with the atomic hopping parameter $t_{\rm a}$ given by
\begin{eqnarray}\label{hopmolec}
t_{{\rm a}} = - \int d{\bf x}~
 w^{{\rm a}*}_{\bf 0}({\bf x} - {\bf x}_{i})
\left[
 - \frac{\hbar^{2} \nabla^{2}}{2 m_{\rm a}}
+V_{0} ({\bf x}) \right] w^{\rm a}_{\bf 0} ({\bf x} - {\bf
x}_{j}),
\end{eqnarray}
where $i$ and $j$ are nearest-neighbouring sites. For the on-site
energy $\epsilon_{\rm a}$ we find
\begin{eqnarray}
\epsilon_{{\rm a}} =
 \int d{\bf x}~
 w^{{\rm a}*}_{\bf 0}({\bf x} - {\bf x}_{i})
\left[
 - \frac{\hbar^{2} \nabla^{2}}{2 m_{\rm a}}
+V_{0} ({\bf x}) \right] w^{\rm a}_{\bf 0} ({\bf x} - {\bf
x}_{i}).
\end{eqnarray}
The hopping parameters $t_{\rm a,m}$ satisfy exactly the relation
\cite{koetsier}
\begin{equation}
t_{\rm a,m}= \frac{\hbar \omega \lambda}{\pi^{3/2}l_{\rm a,m}}~
e^{-(\lambda/\sqrt{2} \pi l_{\rm a,m})^2}
\end{equation}
with $\lambda$ the wavelength of the lattice laser and $l_{\rm
m}=l_{\rm a}/\sqrt{2}$. Furthermore, in the tight-binding limit,
$\epsilon_{\rm a} = 3 \hbar \omega /2$.
\\

Combining equations (\ref{tracef}), (\ref{sigmaf}) and
(\ref{latticesm}), we find that the on-site bare molecular
propagator is given by
\begin{equation}
-\hbar G_{\rm m}^{-1}(i \hbar \omega_{n}) =  - i \hbar \omega_{n}
+ 3 \hbar \omega/2 - 2 \mu + \delta + \hbar \Sigma_m (i
\hbar\omega_{n}).
\end{equation}
with $\delta$ the renormalized detuning (see section
\ref{twobodphys}). We perform an analytic continuation $i \hbar
\omega_{n} \rightarrow E + i 0$ and  the zeros of the above
equation are the poles of the Green's function, which in turn
correspond to the physical eigenstates of the on-site problem. To
study these eigenstates, we calculate the spectral weight function
$\rho(E)$, given by
\begin{equation}
\rho (E) = -\frac{1}{\pi \hbar} {\rm Im} \left[ G_{\rm m}^{(+)}
(E) \right],
\end{equation}
where $G_{\rm m}^{(+)}(E)$ is the retarded Green's function of the
system, i.e., $G_{\rm m}^{(+)}(E) = G_{\rm m} (E + i 0)$. In the
two-body limit the chemical potential is zero and, because the
self-energy is real, the spectral weight function becomes a set of
delta functions located at the solutions $\epsilon_{\sigma}$ of
the equation
\begin{equation} \label{endressfu}
\frac{3}{2} \hbar \omega + \delta = \epsilon_{\sigma}  - g^{2}
\left[ \frac{\Upsilon(\epsilon_{\sigma} - 3 \hbar
\omega/2)}{\sqrt{2} \pi l_{\rm a}^{3} \hbar \omega}  \right],
\end{equation}
where we note that the delta functions have strength $Z_{\sigma}$,
given by
\begin{equation} \label{zsig}
Z_{\sigma} = \left(1- \left. \frac{\partial \hbar \Sigma_{\rm
m}(E)}{\partial E}\right|_{E=\epsilon_{\sigma}}\right)^{-1}.
\end{equation}

We now have all ingredients needed for the effective atom-molecule
theory, since we can substitute $G_{\rm a}^{-1}$ and $G_{\rm
m}^{-1}$ and $g$ in equation (\ref{effact}). In this manner, we
have incorporated the on-site two-body physics exactly in the
effective many-body theory. Furthermore, from the derived
expression for the on-site bare molecular Green's function, we
introduce the convenient notion of dressed molecular fields
$b_{\sigma}$, by making use of
\begin{equation}
-\langle bb^*\rangle \equiv  G_{\rm m}(i \hbar \omega_n)=
\sum_{\sigma} Z_{\sigma} \frac{-\hbar}{-i\hbar \omega_n
+\epsilon_{\sigma}-2\mu} \equiv - \sum_{\sigma} Z_{\sigma }
\langle b_{\sigma}b_{\sigma}^*\rangle.
\end{equation}
Using these dressed molecular fields, we finally find the
Hamiltonian representation of our effective action, given by
\begin{eqnarray} \label{effham}
H &=& -t_{\rm a} \sum_{\langle i,j \rangle} a^{\dagger}_{i}
a^{\phantom \dagger}_{j} -t_{m} \sum_{\sigma} \sum_{\langle i,j
\rangle} b^{\dagger}_{i,\sigma} b^{\phantom \dagger}_{j,\sigma}
\nonumber \\ &&  + \sum_{\sigma} \sum_{i} \left( \epsilon_{\sigma}
- 2 \mu \right) b^{\dagger}_{i,\sigma} b^{\phantom
\dagger}_{i,\sigma}
 + \sum_{i} (\epsilon_{\rm a} -\mu ) a^{\dagger}_{i} a^{\phantom \dagger}_{i}
\nonumber \\ &&  + g' \sum_{\sigma} \sum_{i} \sqrt{Z_{\sigma}}
\left( b^{\dagger}_{i,\sigma} a^{\phantom \dagger}_{i} a^{\phantom
\dagger}_{i} + a^{\dagger}_{i} a^{\dagger}_{i} b^{\phantom
\dagger}_{i,\sigma} \right),
\end{eqnarray}
where $\langle i,j \rangle$ denotes a sum over nearest-neighboring
sites, $a^{\dagger}_{i}$/$a^{\phantom \dagger}_{i}$ denote the
creation/annihilation operators of a single atom at site $i$,
$b^{\dagger}_{i,\sigma}$/$b^{\phantom \dagger}_{i,\sigma}$ denote
the creation/annihilation operators corresponding to the dressed
molecular fields at site $i$, and the effective atom-molecule
coupling in the optical lattice is given by
\begin{equation}
g' = g~ \int d {\bf x} \left[ \phi^{{\rm m}*}_{\bf 0} ({\bf x})
\phi^{\rm a}_{\bf 0} ( {\bf x}) \phi^{\rm a}_{\bf 0} ({\bf x})
\right].
\end{equation}

From equation (\ref{effham}), we see that the effective
Hamiltonian depends on several parameters. In particular, it
depends on the energies of the dressed molecular fields
$\epsilon_{\sigma}$ and the wavefunction renormalization factors
$Z_{\sigma}$. In this section, we calculated these parameters from
first principles by a field-theoretical calculation. But from a
comparison with ref. \cite{feshopt}, we see that exactly the same
results can be obtained from just a two-body calculation on a
single site. This is as expected, since the poles of the molecular
Green's function correspond to the physical eigenstates of the
on-site two-body Feshbach problem.

So far, we did not take into account background atom-atom
scattering, which is expected to be of particular importance for
${}^6$Li, which has an extremely large background scattering
length. Because we use an effective Hamiltonian and dressed
molecular fields, we can include the background scattering in a
very convenient way. Namely, if we are able to solve the two-body
physics on a single site including background scattering, then,
through the notion of our dressed fields, the background
scattering gets automatically incorporated into the many-body
theory. This is what we achieve in the next section.

\section{Two-body physics on a single site.} \label{twobodphys}
Consider the Feshbach problem for two atoms on a single site of an
optical lattice. In the typical case of experimental interest, the
optical lattice potential is deep and the energy of the atoms is
low. As a result, the on-site potential is well approximated by an
isotropic harmonic potential with angular frequency $\omega$.
Furthermore, the atoms interact with each other through the
potential $V_{\rm aa}$ which depends only on their relative
coordinate $\mathbf{r}$, while the atom-molecule coupling is given
by $V_{\rm am}$. The resulting problem can be separated into a
center-of-mass part and a relative part. Only the relative part is
important in solving the two-body physics on a single site, while
the center-of-mass part determines the tunneling of molecules
between adjacent sites.

Taking all this into account, the relative Schr\"{o}dinger
equation for the two-channel Feshbach problem on a single site
becomes
\begin{equation} \label{schreq}
\left( \begin{array}{cc}
H_0 + V_{\rm aa} & V_{\rm am} \\
V_{\rm am} & \delta_{\rm B}
\end{array}
\right) \left( \begin{array}{cc} |\psi_{\rm aa} \rangle \\
|\psi_{\rm m} \rangle \end{array} \right)= E \left( \begin{array}{cc} |\psi_{\rm aa} \rangle \\
|\psi_{\rm m} \rangle \end{array} \right),
\end{equation}
where $\delta_{\rm B}$ denotes the energy of the bare molecular
state and the relative noninteracting atomic Hamiltonian $H_0$ is
given by
\begin{equation}
H_0 = -\frac{\hbar^2 \nabla^2_{\mathbf{r}}}{m_{\rm a}}+
\frac{1}{4} m_{\rm a} \omega^2 \mathbf{r}^2.
\end{equation}
Note that in writing down equation (\ref{schreq}) we have assumed
that the energy of the bare molecular state $|\psi_m \rangle$ is
not affected by the optical lattice. This is well justified, since
the spatial extent of the bare molecular wavefunction, centered
around $\mathbf{r}=0$, is very small compared the the harmonic
oscillator length $l_{\rm a} \equiv \sqrt{\hbar/m_{\rm a}\omega}$.
Furthermore, we can rewrite equation (\ref{schreq}) to obtain the
following equation for the energy eigenvalues
\begin{equation} \label{energyeq}
\langle \psi_{\rm m}| V_{\rm am}\frac{1}{E-H_0-V_{\rm aa}}V_{\rm
am}|\psi_{\rm m} \rangle = E - \delta_{\rm B}.
\end{equation}

\subsection{No background atom-atom scattering.}
For many atoms of interest, such as for example rubidium or
potassium, we have that $|V_{\rm aa}| \ll \hbar \omega$, which
means that we can neglect $V_{\rm aa}$ compared to $H_0$. The
resulting problem was solved in ref. \cite{feshopt} and here we
will highlight the most important results.

The eigenfunctions of $H_0$ can be written in terms of the
generalized Laguerre polynomials
\begin{equation} \label{relwf}
\phi^{\rm aa}_n (r) = \langle \mathbf{r}|\phi^{\rm aa}_n\rangle =
\frac{e^{-r^2/4l_{\rm a}^2} L_n^{1/2}(r^2/2l_{\rm a}^2)}{(2 \pi
l_{\rm a}^2)^{3/4} \sqrt{L_n^{1/2}(0)}},
\end{equation}
which correspond to the eigenenergies
\begin{equation}
E_n = (2n+3/2)\hbar \omega,
\end{equation}
for $ n=0,1,2,...$. Using the completeness relation of the
eigenfunctions $|\phi^{\rm aa}_n \rangle$, equation
(\ref{energyeq}) can be rewritten as
\begin{equation} \label{energysumnoabg}
\sum_{n} \frac{|\langle \psi_{\rm m}|V_{\rm am}|\phi^{\rm
aa}_{n}\rangle|^2}{E-E_n}=E-\delta_{\rm B}.
\end{equation}
Invoking the pseudopotential approxiation for fermions, which
means that $\langle \mathbf{r}|V_{\rm am}| \psi_{\rm m} \rangle =
g \delta(\mathbf{r})$, yields
\begin{eqnarray} \label{buscheq}
E - \delta_{\rm B} &=&g^2 \sum_n \frac{\phi^{{\rm aa}*}_n(0)\phi^{\rm aa}_n(0)}{E-E_n} \nonumber \\
&=& g^2\left[\frac{\Upsilon(E)}{2\sqrt{2}\pi l_{\rm a}^3 \hbar
\omega }-\lim_{r \rightarrow 0} \frac{m}{4 \pi \hbar^2 r} \right],
\end{eqnarray}
where the function $\Upsilon(E)$ was introduced in equation
(\ref{ups}). The energy-independent divergence in equation
(\ref{buscheq}) is an ultraviolet divergence resulting from the
use of pseudopotentials, and was first obtained by Busch \emph{et
al.} in the context of a single-channel problem \cite{busch}. It
can be dealt with by the following renormalization procedure. We
define the renormalized detuning $\delta$ as
\begin{equation} \label{defdet}
\delta \equiv \delta_{\rm B} - \lim_{r \rightarrow 0}\frac{m_{\rm
a}g^2}{4 \pi \hbar^2 r},
\end{equation}
which has two major advantages. Not only do we absorb the
ultraviolet divergence in the definition of $\delta$, but this
renormalized detuning also has a relevant experimental meaning, in
contrast to $\delta_{\rm B}$. Namely, $\delta$ corresponds to the
detuning from the magnetic field $B_0$, at which the Feshbach
resonance takes place in the absence of an optical lattice. This
can be understood from the treatment of the homogeneous Feshbach
problem without an optical lattice \cite{homog}. Here, the
condition for the location of the Feshbach resonance is that the
dressed molecular energy is equal to the threshold of the atomic
continuum. This leads to the resonance condition $\delta_B =
\lim_{r \rightarrow 0}m_{\rm a}g^2/4 \pi \hbar^2 r$. As a result,
the definition in equation (\ref{defdet}) places the resonance in
the absence of an optical lattice conveniently at $\delta=0$ and,
by construction, the (renormalized) detuning is of the following
form
\begin{equation}
\delta(B)=\Delta \mu (B-B_0),
\end{equation}
where $\Delta \mu$ is the difference in magnetic moment between
the atoms in the open channel and the bare molecule in the closed
channel.

Substituting equation (\ref{defdet}) in equation (\ref{buscheq})
gives
\begin{equation} \label{eneqnoabgdis}
E - \delta = \frac{g^2 \Upsilon(E)}{2\sqrt{2}\pi l_{\rm a}^3 \hbar
\omega},
\end{equation}
which allows us to calculate the eigenenergies of the Feshbach
problem on a single site given a certain detuning $\delta$. Note
the similarity between  equation (\ref{eneqnoabgdis}) and equation
(\ref{endressfu}). The only two differences are a constant shift
in the energy of $3 \hbar \omega/2$ and a factor of 2, which are
both readily explained. The shift of $3 \hbar \omega/2$ is just
the center-of-mass energy neglected in this relative calculation.
The factor of 2 is due to the fact that here we consider fermions,
while in the previous section we considered bosons. Since
equations (\ref{endressfu}) and (\ref{eneqnoabgdis}) are for the
rest completely equivalent, we see that indeed the energies
$\epsilon_{\sigma}$ of the dressed molecular fields $b_{\sigma}$,
needed for the effective Hamiltonian of equation (\ref{effham}),
can just as well be determined from a two-body calculation, as
from a field-theoretical calculation. An obvious advantage of the
two-body approach is that it is much simpler. Furthermore, as we
show in the next paragraph, the two-body approach also allows for
an incorporation of atom-atom scattering effects, which is of
course also possible, but much more difficult from a
field-theoretical point of view. These effects are particularly
important for the fermion ${}^6$Li, which has a very large
background atom-atom scattering length $a_{\rm bg}$ near the
extremely broad Feshbach resonance at $B_0 = 834$ G, namely on the
order of $-1500 a_0$, where $a_0$ is the Bohr radius.

\subsection{With background atom-atom interaction.}
The relative Schr\"{o}dinger equation for two atoms in a harmonic
potential interacting through the pseudopential $V_{\rm aa}=V_0
\delta(\mathbf{r})$ without coupling to a molecular state
\begin{equation} \label{schreqat}
(H_0 + V_{\rm aa}) |\phi_{\nu}\rangle = E_{\nu} |\phi_{\nu}\rangle
\end{equation}
can be solved analytically for $s$-wave scattering following a
treatment along the lines of ref. \cite{busch}.  We just note that
in this reference a suspicious `molecular' bound state appears,
which we believe to be unphysical for several reasons. Most
convincing is probably to consider the limit $a_{\rm bg}
\rightarrow 0^{+}$, in which case the interaction vanishes and we
should recover the case of a simple harmonic oscillator. This is
indeed what happens in ref. \cite{busch}, except for the
suspicious bound state, whose energy goes to minus infinity. Since
such a state does not arise in the treatment of the harmonic
oscillator, we conclude that the state is unphysical and we will
exclude it from further calculations. Note also that for the
specific case of ${}^6$Li this issue is not relevant, since the
suspicious state only occurs for positive values of $a_{\rm bg}$.

The eigenvalues of equation (\ref{schreqat}) are given by
$E_{\nu}=(2 \nu + 3/2)\hbar \omega$, which are the solutions to
the equation
\begin{equation} \label{abg}
\sqrt{2}\frac{\Gamma(-E_{\nu}/2\hbar\omega
+3/4)}{\Gamma(-E_{\nu}/2\hbar\omega +1/4)}= \frac{l_{\rm
a}}{a_{\rm bg}},
\end{equation}
where $\nu$ is in general not an integer. The corresponding
eigenfunctions $\phi^{\rm aa}_{\nu}(r)=\langle r|\phi^{\rm
aa}_{\nu}\rangle$ have the following form
\begin{equation}
\phi^{\rm aa}_{\nu}(r)=A_{\nu} e^{-r^2/4l_{\rm a}^2}
\Gamma(-\nu)U(-\nu,3/2,r^2/2l_{\rm a}^2),
\end{equation}
where $A_{\nu}$ is a normalization constant and
$U(-\nu,3/2,r^2/2l^2)$ is a so-called confluent hypergeometric
function of the second kind. We can determine $A_{\nu}$ by
applying a limiting procedure to theorem 7.622 of ref.
\cite{gradrys}, resulting in
\begin{equation} \label{intcst}
A_{\nu}^2=\frac{\Gamma(-\nu-1/2)}{4 \sqrt{2} \pi^2l_{\rm a}^3
\Gamma(-\nu)[\psi^0(-\nu)-\psi^0(-\nu-1/2)]},
\end{equation}
in which $\psi^0$ denotes the digamma function. Furthermore, we
can use the completeness relation of the eigenfunctions
$|\phi^{\rm aa}_{\nu}\rangle$ to rewrite equation
(\ref{energyeq}), which yields
\begin{equation} \label{energysum}
\sum_{\nu} \frac{|\langle \psi_{\rm m}|V_{\rm am}|\phi^{\rm
aa}_{\nu}\rangle|^2}{E-E_{\nu}}=E-\delta_{\rm B}.
\end{equation}

\begin{figure}[t]
\centering
\includegraphics[width=1.0\textwidth]{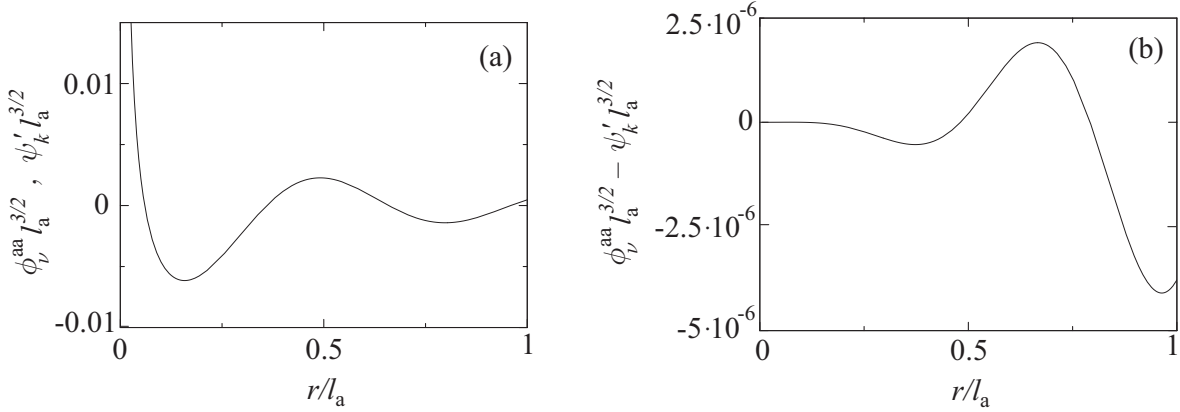}
\caption{(a) Plot of the wavefunctions $\phi^{\rm aa}_{\nu}$ and
$\psi'_{k}$ as a function of the interatomic distance $r$. On the
scale of the figure, the two wavefunctions cannot be
distuinguished. (b) Plot of the difference between the
wavefunctions $\phi^{\rm aa}_{\nu}$ and $\psi'_{k}$ as a function
of the interatomic distance $r$. Note that this difference is
extremely small compared to the value of the wavefunctions
themselves. All quantities on the axes are made dimensionless
using the atomic harmonic oscillator length $l_{\rm a}$
\label{wavefu}}.
\end{figure}

In order to proceed from equation (\ref{energysum}), we realize
that the integral $\langle \psi_{\rm m}|V_{\rm am}|\phi^{\rm
aa}_{\nu}\rangle$ only acquires a finite contribution from a very
small region around $\mathbf{r}=0$, because the spatial extent of
both the molecular wavefunction $\psi_{\rm m}$ and the
atom-molecule coupling $V_{\rm am}$ is very small. This means that
we are only interested in $\phi^{\rm aa}_{\nu}(r)$ for small
values of $r$, giving
\begin{eqnarray}
\phi^{\rm aa}_{\nu}(r)&=&
-\sqrt{\pi}A_{\nu}\left(\frac{2\Gamma(-\nu)}{\Gamma(-\nu-1/2)}-\sqrt{2}
\frac{l_{\rm a}}{r}+\mathcal{O}\left(\frac{r}{l_{\rm a}}\right) \right) \nonumber \\
&=& -\sqrt{2 \pi}l_{\rm a}A_{\nu}\left( \frac{1}{a_{\rm bg}}
-\frac{1}{r}+ \mathcal{O}\left(\frac{r}{l^2_{\rm
a}}\right)\right),
\end{eqnarray}
where in the second line we used equation (\ref{abg}). Note that
the eigenfunctions $\phi^{\rm aa}_{\nu} (r)$ behave for small $r$
completely analogous to the $s$-wave scattering states in the
absence of an optical lattice $\psi_k (r)$, which are of the
following form
\begin{equation}
 \psi_k (r) = \frac{\sin\left[kr +
\delta_0(k)\right]}{kr},
\end{equation}
where $k$ is the relative momentum of the scattering state and
$\delta_0(k)$ is the so-called $s$-wave phase shift, given by
\begin{eqnarray}
k &=& \sqrt{\frac{m_{\rm a} E_{\nu}}{\hbar^2}},  \\
\delta_0(k)&=&\tan^{-1}(-k a_{\rm bg}). \label{phaseshift}
\end{eqnarray}
Indeed, for small $r$, we have
\begin{eqnarray}
\psi_k (r)&=&\cos \delta_0(k)+ \frac{\sin \delta_0(k)}{kr} +
\mathcal{O}(kr) \nonumber \\
&=& -\frac{\sin \delta_0(k)}{k}\left( \frac{1}{a_{\rm bg}} -
\frac{1}{r} +\mathcal{O}(k^2r) \right),
\end{eqnarray}
where in the second line we used equation (\ref{phaseshift}). As a
result, for small $r$
\begin{equation} \label{phipsi}
 \phi^{\rm aa}_{\nu}(r) = \sqrt{2 \pi}\frac{k l_{\rm a}  A_{\nu}}{\sin \delta_0(k)}\psi_k
 (r).
\end{equation}
This relation is illustrated by figure \ref{wavefu}, where
$\phi^{\rm aa}_{\nu}$, $\psi_k' \equiv \sqrt{2 \pi}k l_{\rm a}
A_{\nu}\psi_k/\sin \delta_0$  and their difference are plotted as
a function of $r$. We see that the difference indeed vanishes for
small $r$ and increases with increasing $r$. The scattering states
$\psi_k (r)$ are the solutions of the relative two-atom $s$-wave
scattering problem without optical lattice and without
atom-molecule coupling \cite{homog}. The physical reason for the
similarity between $\phi_{\nu}(r)$ and $\psi_k(r)$ near $r=0$ is
due to the following: for small $r$ compared to $l_{\rm a}$, the
atoms experience an effectively constant harmonic potential by
which they are not affected. Therefore, for small $r$ the
wavefunction from the theory with optical lattice should, up to a
normalization, reduce to the wavefunction from the theory without
optical lattice. Since it is known how to solve the homogeneous
Feshbach problem without an optical lattice \cite{homog}, we are
able to profit from this knowledge. This is what we do in the next
paragraph.

\subsection{Solving the two-atom Feshbach-problem.}
By combining equations (\ref{energysum}) and (\ref{phipsi}), we
obtain
\begin{equation} \label{energysum2}
\sum_{\nu} \frac{2 \pi k^2 l_{\rm a}^2 A_{\nu}^2}{\sin^2
\delta_0(k)} \frac{|\langle \psi_{\rm m}|V_{\rm
am}|\psi_{k}\rangle|^2}{E-E_{\nu}}=E-\delta_{\rm B}.
\end{equation}
From ref. \cite{homog}, we have that
\begin{equation} \label{coupling}
\langle \psi_{\rm m}|V_{\rm am}|\psi_{k}\rangle =
\frac{g}{1-ia_{\rm bg}k}.
\end{equation}
Substituting equations (\ref{intcst}) and (\ref{coupling}) into
equation (\ref{energysum2}) and rewriting the result with the use
of equation (\ref{phaseshift}) yields
\begin{equation} \label{energysum3}
\frac{g^2}{2 \pi a_{\rm bg}l_{\rm a}^2}\sum_{\nu} \frac{1}{\left[
E-(2\nu+3/2) \right]\left[ \psi^0(-\nu)-\psi^0(-\nu-1/2) \right]}
=E - \delta_{\rm B}.
\end{equation}

Just like in section 3.1, we would like to renormalize the bare
detuning $\delta_{\rm B}$, which has no experimental meaning, to
the renormalized detuning $\delta$, which gives the detuning from
the magnetic field $B_0$ at which the Feshbach resonance takes
place in the absence of an optical lattice. From the homogeneous
theory that takes $a_{\rm bg}$ into account \cite{homog}, we find
at resonance that
\begin{equation}
\delta_{\rm B} = \frac{g^2 m_{\rm a}}{4 \pi \hbar^2 |a_{\rm bg}|}.
\end{equation}
Defining the renormalized detuning $\delta$ as $\delta \equiv
\delta_B - g^2 m_{\rm a}/4 \pi \hbar^2 |a_{\rm bg}|$ places the
Feshbach resonance in the absence of an optical lattice
conveniently at $\delta=0$ and brings equation (\ref{energysum3})
into its final form
\begin{eqnarray} \label{energysumfin}
E - \delta&=&\frac{g^2}{2 \pi a_{\rm bg}l_{\rm a}^2}\sum_{\nu}
\frac{1}{\left[ E-(2\nu+3/2) \right]\left[
\psi^0(-\nu)-\psi^0(-\nu-1/2) \right]}
\nonumber \\
&&+ \frac{g^2 m_{\rm a}}{4 \pi \hbar^2 |a_{\rm bg}|},
\end{eqnarray}
where by construction $\delta= \Delta \mu(B-B_0)$ again. As in
section 3.1, the term on the right-hand side of equation
(\ref{energysumfin}) can be interpreted as a molecular self-energy
$\hbar \Sigma_{\rm m}(E)$. Furthermore, upon going to a many-body
theory, the solutions of equation (\ref{energysumfin}) can be
interpreted as the energies $\epsilon_{\sigma}$ of the dressed
molecular fields $b_{\sigma}$, resulting in a generalized Hubbard
Hamiltonian in the same way as before.

For ${}^6$Li, which has a very large $a_{\rm bg}$, we expect the
effects of the background scattering to be of particular
importance. Therefore, in the next paragraph, we study equation
(\ref{energysumfin}) for the specific case of ${}^6$Li to see if
it leads to significant differences from the results given by
equation (\ref{eneqnoabgdis}), which doesn't take background
scattering into account.

\subsection{Lithium}
\begin{figure}[t]
\centering
\includegraphics[width=0.9\textwidth]{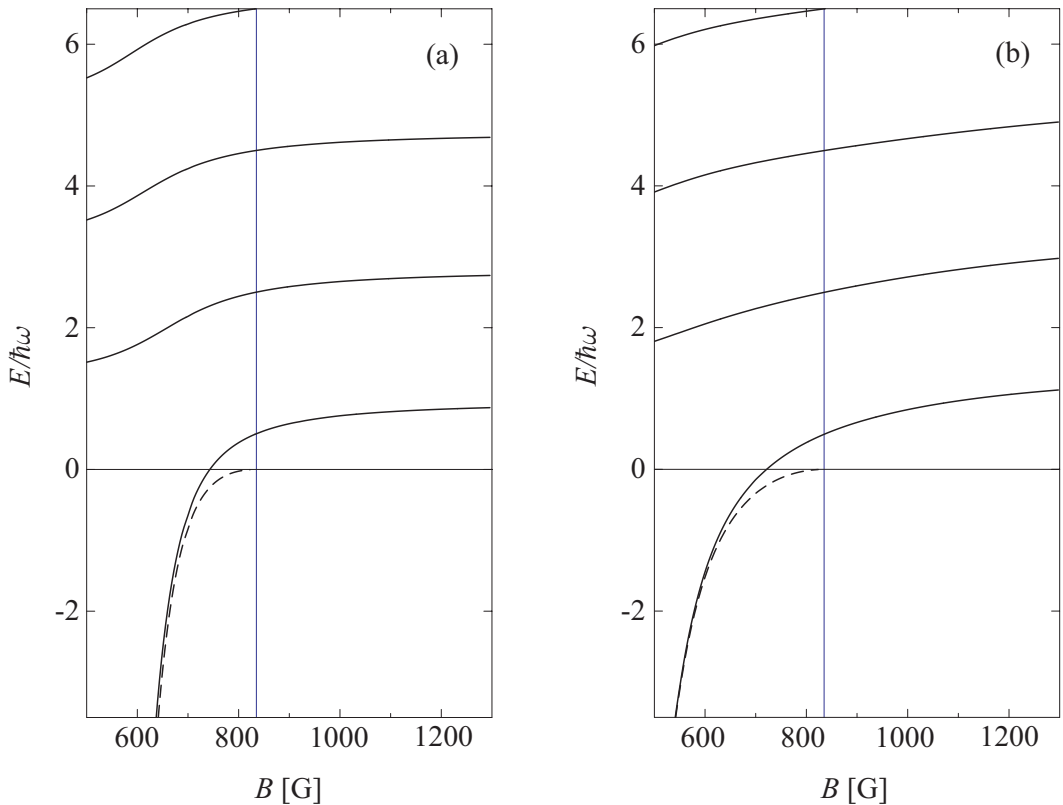}
\caption{(a) Relative energy levels (solid lines) as a function of
the magnetic field $B$ for a system consisting of two ${}^6$Li
atoms in the hyperfine states $|1 \rangle$ and $|2 \rangle$ near
the Feshbach resonance at 834 G in a harmonic potential of $\omega
= 10^6$ ${\rm s}^{-1}$ with background atom-atom scattering taken
into account. The dashed line corresponds to the binding energy of
a dressed ${}^6$Li$_2$-molecule in the absence of an optical
lattice. Note that as the eigenenergy of the dressed molecular
ground state with optical lattice (solid line) decreases, it
converges to the dressed molecular eigenenergy without optical
lattice (dashed line). (b) Similar to panel (a), only this time
the background atom-atom scattering is not taken into account.
Note that as a result both panels differ significantly.
\label{lithenergy}}
\end{figure}

Figure \ref{lithenergy}a shows the energy-levels of two
interacting ${}^6$Li atoms in the hyperfine states $|1 \rangle$
and $|2 \rangle$ on a single site near the Feshbach resonance at
$834$ G. The diagram has been obtained by numerically solving
equation (\ref{energysumfin}), where we used the known
experimental values for $g(B)$ and $a_{\rm bg}(B)$ corresponding
to ${}^6$Li \cite{falco}. Note that for the extremely broad
Feshbach resonance at $834$ G these parameters depend on the
magnetic field $B$. Furthermore, we took for $l_{\rm a}$ the
realistic value $10^{-7}$ m, corresponding to $\omega = 10^6$
${\rm s}^{-1}$. In figure \ref{lithenergy}b the same diagram is
shown, but this time without taking the background atom-atom
scattering into account. This diagram has been obtained by
numerically solving equation (\ref{eneqnoabgdis}), where again the
experimental value of the parameter $g(B)$, corresponding to
${}^6$Li, was substituted.

From figure \ref{lithenergy}, we see that for ${}^6$Li near the
Feshbach resonance at 834 G it is important to include background
atom-atom interactions, since it leads to a significant adjustment
of the corresponding two-body energy-level diagram. Upon inclusion
of the background atom-atom interactions, the horizontal
asymptotes are shifted to different energy values and the avoided
crossings, see also ref. \cite{feshopt}, become much less broad.
Physically, this is a result of the fact that the large value of
$a_{\rm bg}$ effectively reduces the atom-molecule coupling.

\begin{figure}[t]
\centering
\includegraphics[width=1.0\textwidth]{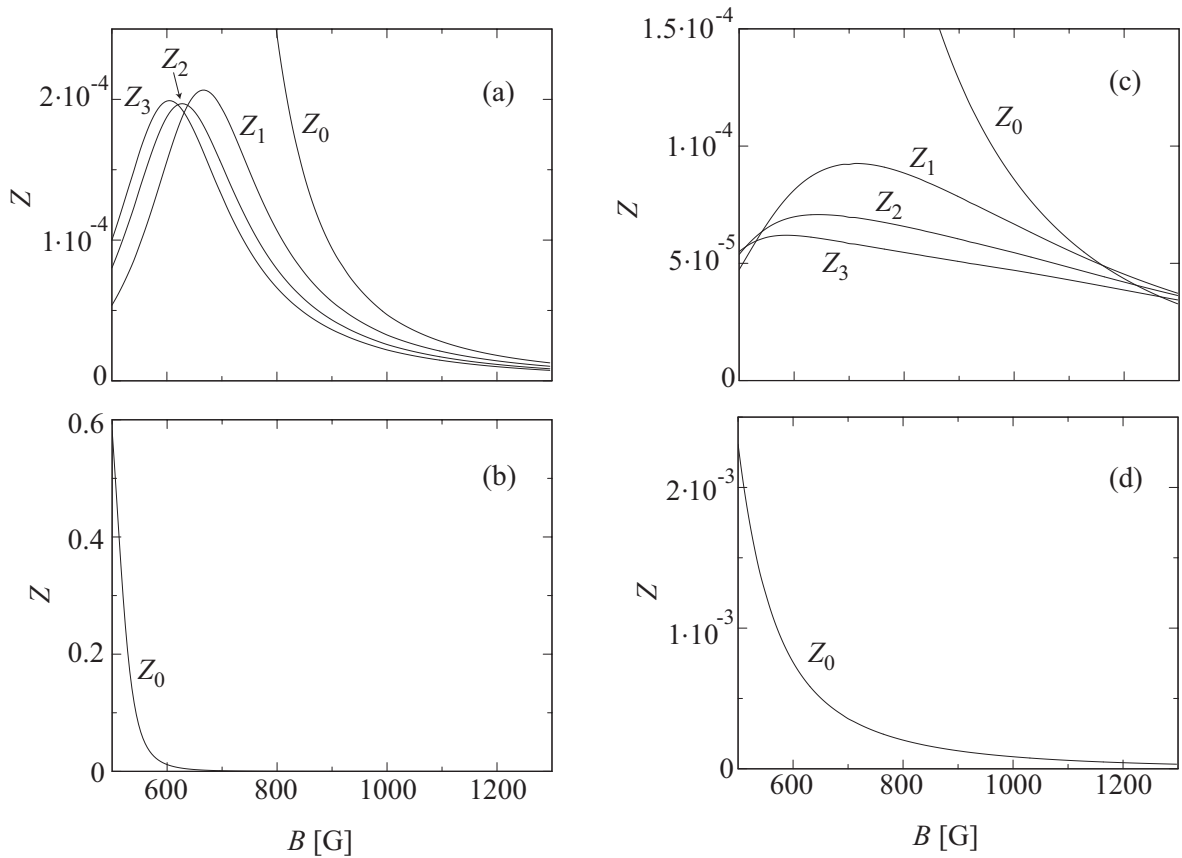}
\caption{(a),(b) Wavefunction renormalization factors $Z_{\sigma}$
for ${}^6$Li near the Feshbach resonance at 834 G as a function of
the magnetic field $B$ with background atom-atom scattering taken
into account. $Z_{0}$ is the renormalization factor corresponding
to the ground state of the two-body on-site Feshbach problem,
$Z_{1}$ is the renormalization factor corresponding to the first
excited state, and so on. (c),(d) Wavefunction renormalization
factors $Z_{\sigma}$ for ${}^6$Li as a function of the magnetic
field $B$ without background atom-atom scattering. Note that the
panels on the left differ significantly from the panels on the
right. \label{lithZ}}
\end{figure}

The difference between equations (\ref{eneqnoabgdis}) and
(\ref{energysumfin}) in the case of ${}^6$Li becomes even clearer
when we look at the corresponding wavefunction renormalization
factors $Z_{\sigma}$ given by equation (\ref{zsig}), which are
needed for the construction of the effective Hamiltonian given by
equation (\ref{effham}). For the calculation of $Z_{\sigma}$ we
need the molecular self-energy $\hbar \Sigma_{\rm m}$, which is
given by the right-hand side of equation (\ref{eneqnoabgdis}) for
the case without background atom-atom scattering and by the
right-hand side of equation (\ref{energysumfin}) for the case with
background atom-atom scattering. In figure \ref{lithZ}, the
corresponding wavefunction renormalization factors $Z_{\sigma}$
are plotted for both cases as a function of the magnetic field
$B$. Here, $Z_0$ corresponds to the groundstate of the relative
on-site two-body Hamiltonian, $Z_1$ corresponds to the first
excited state, and so on. Note that the two-body renormalization
factors $Z_{\sigma}$ obtained including background interactions
(figure \ref{lithZ}a), are very different from the factors
obtained without background interactions (figure \ref{lithZ}b).
Also note that these factors are directly experimentally
observable as demonstrated by Partridge {\it et al.}
\cite{cross6}. We thus conclude that background atom-atom
scattering plays a significant role for an atomic gas of ${}^6$Li
atoms in an optical lattice near the Feshbach resonance at 834 G.

\section{Conclusion}
In summary, we have derived from first principles an effective
atom-molecule theory that describes an atomic gas in an optical
lattice near a Feshbach resonance. The theory was formulated in
the tight-binding limit and incorporated all the relevant two-body
physics exactly. The field-theoretical derivation reconfirmed that
an atomic gas in an optical lattice near a Feshbach resonance is
accurately described by a generalized Hubbard model, as first
obtained in ref. \cite{feshopt}. In the original approach of ref.
\cite{feshopt} the generalized Hubbard model was derived in an
easier way by starting from a two-body calculation on a single
site. Due to the equivalence of both approaches, we used the
latter to formulate a way in which to include the background
atom-atom scattering into the many-body theory, which has not been
done previously. To this end, we showed how to solve exactly the
two-channel Feshbach problem for two atoms on a single site
including background atom-atom scattering. The solution was
applied to ${}^6$Li near the experimentally relevant Feshbach
resonance at 834 G. Specifically, the two-body on-site energy
levels and the wavefunction renormalization factors $Z_{\sigma}$
were obtained, which are needed for the generalized Hubbard
Hamiltonian describing the many-body physics. As it turned out,
the various energy levels and the wavefunction renormalization
factors obtained with background interactions taken into account,
are significantly different from the energy levels and the
renormalization factors obtained without background interactions
taken into account. From this we conclude that in the case of
${}^6$Li background atom-atom scattering plays an important role
and cannot be neglected for an accurate microscopic description of
future experiments with atomic lithium in an optical lattice.

\end{document}